\documentclass{article}%
\usepackage{amsmath}
\usepackage{amsfonts}
\usepackage{amssymb}
\usepackage{graphicx}%
\providecommand{\U}[1]{\protect\rule{.1in}{.1in}}

\begin{document}

\author{Giuseppe Castagnoli\\\textit{Elsag Bailey ICT Division and Quantum Information Laboratory}\\\textit{Via Puccini 2, 16154 Genova, Italy }}
\title{Completing the physical representation of quantum algorithms provides a
retrocausal explanation of their speedup}
\maketitle

\begin{abstract}
In previous works, we showed that an optimal quantum algorithm can always be
seen as a sum over classical histories in each of which the problem solver
knows in advance one of the possible halves of the solution she will read in
the future and performs the computation steps (oracle queries) still needed to
reach it. Given an oracle problem, this retrocausal explanation of the speedup
yields the order of magnitude of the number of oracle queries needed to solve
it in an optimal quantum way. Presently, we provide a fundamental
justification for the explanation in question and show that it comes out by
just completing the physical representation of quantum algorithms. Since the
use of retrocausality in quantum mechanics is controversial, showing that it
answers the well accepted requirement of the completeness of the physical
description should be an important pass.

\end{abstract}

\section{Foreword}

The \textit{quantum computational speedup} is the fact that quantum algorithms
solve the respective problems with fewer computation steps (oracle queries)
than their best classical counterparts, sometimes demonstrably fewer than
classically possible.

A paradigmatic example is the simplest instance of the quantum algorithm
devised by Grover $\left[  1\right]  $. Bob, the problem setter, hides a ball
in one of four drawers. Alice, the problem solver, is to locate it by opening
drawers. In the classical case, Alice has to open up to three drawers, always
one in the quantum case (the problem is an example of \textit{oracle problem}
and the operation of checking whether the ball is in a drawer is an example of
\textit{oracle query}).

Deutsch\textit{ }$\left[  2\right]  $ commented his 1985 discovery of the
seminal quantum speedup, of course allowed by quantum superposition and
interference, with the statement that \textit{computation is physical}%
\footnote{In 1982, Feynman had pointed out that the classical simulation of a
quantum process can require an amount of time $\times$\ physical resources
exponentially higher than those involved in the quantum process itself
$\left[  3\right]  $.}. This is as simple as deep. The interplay between
computing as a mental process and some "outside" physical process, like
counting on the fingers, must be as ancient as the idea of computation itself.
However, until the physical process has remained classical, its character has
not enriched the idea of computation. The turning point comes with quantum
computation. This time physical computation is richer than our idea of
computation, as a matter of fact in ways that have yet to be well understood.

It should be noted that each quantum algorithm has been found by means of
ingenuity. In mainstream literature, there is no fundamental explanation of
the speedup, no unification of the various speedups (quadratic, exponential),
no way of foreseeing the number of oracle queries needed to solve a generic
oracle problem in an optimal quantum way.

Here we ascribe our limited understanding of the speedup to the fact that the
physical representation of quantum algorithms arrived half done. Being limited
to the input-output transformation typical of computation, it is physically
incomplete. It consists of a unitary transformation followed by the final
measurement required to read the solution. As well known, the complete
representation of a quantum process must include the initial measurement, the
unitary transformation of the state after measurement, and the final measurement.

Preliminary versions of the present retrocausal explanation of the speedup
were already provided in the evolutionary approach $\left[  4\div9\right]  $.
In the present work we show that just completing the physical representation
of quantum algorithms provides the explanation in question. This also
clarifies the roles that time-symmetric and relational quantum mechanics play
in it.

\section{Extended summary}

We are in the context of \textit{oracle computing}. Let $\left\{
f_{\mathbf{b}}\left(  \mathbf{a}\right)  \right\}  $ be a set of functions,
with $\mathbf{b}$\ the function identifier and $\mathbf{a}$\ the argument of
the function. Bob chooses one of these functions -- i. e. a value of
$\mathbf{b}$\ -- and gives Alice the black box that computes it. Alice knows
the set of functions but not Bob's choice. To her the function identifier
$\mathbf{b}$ -- from now on\textit{ the problem setting} -- is hidden inside
the black box. She is to find some characteristic of the function computed by
the black box (e. g. the period in the case of a set of periodic functions) by
performing \textit{function evaluations} for appropriate values of
$\mathbf{a}$. By the way, in literature, the black box is also called
\textit{oracle} and function evaluation \textit{oracle query}.

The usual representation of quantum algorithms, limited to Alice's
problem-solving action, consists of the input-output transformation typical of
computations. Alice works on a quantum register $A$. The input state of this
register is a pure quantum state independent of the value of $\mathbf{b}$
chosen by Bob. By performing function evaluations interleaved with other
suitable transformations, she unitarily sends the input state into an output
state where register $A$ contains the solution of the problem, namely the
characteristic of the function computed by the black box; in this summary we
assume for simplicity that the solution, $s\left(  \mathbf{b}\right)  $, is a
one to one function of $\mathbf{b}$. Then she acquires the solution by
measuring the content of $A$. This representation has thus the form:%

\begin{equation}%
\begin{tabular}
[c]{lll}%
$\left\vert \mathbf{0}\right\rangle _{A}\overset{U_{A}}{\rightarrow}$ &
$\left\vert s\left(  \mathbf{b}_{c}\right)  \right\rangle _{A}\overset{M_{A}%
}{\rightarrow}$ & $\left\vert s\left(  \mathbf{b}_{c}\right)  \right\rangle
_{A}$\\
&  & \\
$t_{1}$ & $t_{2}$ & $t_{2+}$%
\end{tabular}
\label{usual}%
\end{equation}
In the input state of the quantum algorithm at time $t_{1}$, register $A$
contains a bit string of all zeroes -- but any pure quantum state would do.
$U_{A}$ is the unitary part of Alice's action, it sends $\left\vert
\mathbf{0}\right\rangle _{A}$ into an output state where register $A$ contains
the solution of the problem $s\left(  \mathbf{b}_{c}\right)  $ --
$\mathbf{b}_{c}$\ being the value of $\mathbf{b}$\ chosen by Bob. $M_{A}$ is
the measurement of the content of register $A$; the states before and after
this measurement are the same; we call $t_{i+}$ the time of the state
immediately after a measurement performed at time $t_{i}$.

This representation is physically incomplete in two related ways

(i) It lacks the initial measurement, whereas the complete representation of a
quantum process must consist of an initial measurement, a unitary
transformation of the measurement outcome, and a final measurement.

(ii) There is no physical representation of the value of $\mathbf{b}$ chosen
by Bob. This value is of course essential in determining the quantum process:
it determines the result of function evaluations.

We restore both the initial measurement and the physical representation of the
value of $\mathbf{b}$ by extending the representation of the quantum algorithm
to the process of setting the problem.

We add a possibly imaginary register $B$, under the control of Bob, that
contains $\mathbf{b}$. We assume that, initially, this register is in a
maximally mixed state. Bob measures the content of this register selecting a
problem setting, a value of \ $\mathbf{b}$, at random. Then he unitarily
changes the state after measurement into the state that encodes the desired
problem setting. For simplicity, for the time being, we jump the
transformation in question -- we assume that Bob chooses the random outcome of
the initial measurement.

The representation of the quantum algorithm is now:%

\begin{equation}%
\begin{tabular}
[c]{llll}%
$\frac{1}{\sqrt{c_{\mathbf{\sigma}}}}%
{\textstyle\sum\limits_{\mathbf{b\in\sigma}}}
\operatorname{e}^{i\varphi_{\mathbf{b}}}\left\vert \mathbf{b}\right\rangle
_{B}\left\vert \mathbf{0}\right\rangle _{A}\overset{M_{B}}{\rightarrow}$ &
$\left\vert \mathbf{b}_{c}\right\rangle _{B}\left\vert \mathbf{0}\right\rangle
_{A}\overset{U_{A}}{\rightarrow}$ & $\left\vert \mathbf{b}_{c}\right\rangle
_{B}\left\vert s\left(  \mathbf{b}_{c}\right)  \right\rangle _{A}%
\overset{M_{A}}{\rightarrow}$ & $\left\vert \mathbf{b}_{c}\right\rangle
_{B}\left\vert s\left(  \mathbf{b}_{c}\right)  \right\rangle _{A}$\\
&  &  & \\
$t_{1}$ & $t_{1+}$ & $t_{2}$ & $t_{2+}$%
\end{tabular}
\label{bob}%
\end{equation}
To keep the usual ket vector representation of quantum algorithms, the
maximally mixed state of register $B$\ is represented as a dephased quantum
superposition of all the possible values of $\mathbf{b}$; $\mathbf{b}$ ranges
over the set of all the possible problem settings $\sigma$, of cardinality
$c_{\sigma}$; the $\varphi_{i}$ are independent completely random phases (see
what follows).\ $M_{B}$ is the measurement of the content of register $B$,
which selects a problem setting -- here $\mathbf{b}_{c}$ -- at random (note
that this measurement commutes with that of the content of $A$). At time
$t_{1+}$ we have thus the new input state of the quantum algorithm; from here
onwards the quantum states are those of the usual quantum algorithm but for
the multiplication by $\left\vert \mathbf{b}_{c}\right\rangle _{B}$.

We note that we are making a trivial use of the present\textit{ random phase
representation }$\left[  10\right]  $ of a maximally mixed state: we can
always ignore the random character of the $\varphi_{i}$ but for the
computation of the entropy of the quantum state, which is $n$\ bit, with
$n=\lg_{2}c_{\sigma}$.

This extended representation is very similar to the usual one. However, there
is an important consequence. Before going to it, it is convenient to make a
step backward to explicit things that are usually given for understood.

As well known, the quantum state encapsulates everything that can be known
about the quantum system between observations. To present ends, the question
"known by whom?" is essential. We adopt the answer of the Copenhagen
interpretation: known by the observer.

We still need to clarify who is the observer. The extended representation is
certainly the representation to Bob, the problem setter, and any external
observer: it tells all of them that the problem setting is completely
undetermined at time $t_{1}$, is $\mathbf{b}_{c}$ at time $t_{1+}$, etc. The
point is that it cannot be the representation to Alice, the problem solver.
The sharp state of register $B$\ at time $t_{1+}$ would tell her that the
problem setting chosen by Bob is $\mathbf{b}_{c}$ (see the second state from
the left of array \ref{bob}). It would tell her the solution of the problem
before she begins her problem solving action: in fact we must think that the
function $s\left(  \mathbf{b}\right)  $ is known to Alice.

As is well known, to Alice, the value of $\mathbf{b}$ chosen by Bob must be
hidden inside the black box. This is why the box is called black in classical
computer science. However, while in the classical case it suffices to keep
this concealment in mind, to do the right thing at the right moment, now we
are in the quantum case where, according to the present objective, all
computational facts should be physical.

To represent the concealment physically, we must resort to the relational
interpretation of quantum mechanics $\left[  11,12\right]  $. According to it,
a quantum state has meaning with respect to an observer, like in the
Copenhagen interpretation. What the relational interpretation rejects is the
notion of absolute, or observer-independent, state of a system\textit{.} In
equivalent terms, it rejects the notion of observer-independent values of
physical quantities\textit{ }$\left[  12\right]  $. This notion "\textit{would
be inadequate to describe the physical world beyond the }$\hbar\rightarrow
0$\textit{ limit, in the same way in which the notion of observer-independent
time is inadequate to describe the physical world beyond the }$c\rightarrow
\infty$\textit{ limit}"\textit{ }$\left[  12\right]  $.

In the representation to Alice, we postpone the projection of the quantum
state due to the initial measurement until the end of the unitary part of her
action. As is well known, the projection due to a quantum measurement can be
postponed at will along a unitary transformation that follows it. The
representation becomes:%

\begin{equation}%
\begin{tabular}
[c]{llll}%
$\frac{1}{\sqrt{c_{\mathbf{\sigma}}}}%
{\textstyle\sum\limits_{\mathbf{b\in}\sigma}}
\operatorname{e}^{i\varphi_{\mathbf{b}}}\left\vert \mathbf{b}\right\rangle
_{B}\left\vert \mathbf{0}\right\rangle _{A}\overset{M_{B}}{\rightarrow}$ &
$\frac{1}{\sqrt{c_{\mathbf{\sigma}}}}%
{\textstyle\sum\limits_{\mathbf{b\in}\sigma}}
\operatorname{e}^{i\varphi_{\mathbf{b}}}\left\vert \mathbf{b}\right\rangle
_{B}\left\vert \mathbf{0}\right\rangle _{A}\overset{U_{A}}{\rightarrow}$ &
$\frac{1}{\sqrt{c_{\mathbf{\sigma}}}}%
{\textstyle\sum\limits_{\mathbf{b\in}\sigma}}
\operatorname{e}^{i\varphi_{\mathbf{b}}}\left\vert \mathbf{b}\right\rangle
_{B}\left\vert s\left(  \mathbf{b}\right)  \right\rangle _{A}\overset{M_{A}%
}{\rightarrow}$ & $\left\vert \mathbf{b}_{c}\right\rangle _{B}\left\vert
s\left(  \mathbf{b}_{c}\right)  \right\rangle _{A}.$\\
&  &  & \\
$t_{1}$ & $~t_{1+}$ & $t_{2}$ & $t_{2+}$%
\end{tabular}
\label{alice}%
\end{equation}
Now the maximally mixed state of register $B$ remains unaltered throughout
$M_{B}$, as the associated projection of the quantum state is postponed. The
$n$\ bit entropy of the state of register $B$ in the input state of the
quantum algorithm (at time $t_{1+}$) represents Alice's complete ignorance of
the problem setting chosen by Bob. $U_{A}$ sends this input state into a
mixture of tensor products, each the product of one of the possible problem
settings and the corresponding solution (it maximally entangles the contents
of registers $B$ and $A$). Eventually Alice measures the content of $A$,
selecting the solution corresponding to the problem setting chosen by Bob. The
measurement outcome cannot be predicted by Alice as usual, it is already known
to Bob and any external observer.

Until now, the process of completing the physical representation of the
quantum algorithm brought us to two time-symmetric and relational
representations, one with respect to Bob and any external observer, the other
with respect to Alice.

In either representation there is a unitary transformation between the initial
and final measurement outcomes. There is a consequent ambiguity. The selection
(determination) of the random outcome of the initial measurement and the
corresponding outcome of the final measurement-- of the pair $\mathbf{b}_{c}$
and $s\left(  \mathbf{b}_{c}\right)  $\ -- can be ascribed indifferently to
the initial measurement of the content of $B$ or the final measurement of the
content of $A$.

In the latter case, we should think that the projection of the quantum state
associated with the initial measurement is removed. As a consequence, in
either representation, the projection of the quantum state associated with the
final measurement of the content of $A$ becomes that of the representation to
Alice, namely:%
\[
\frac{1}{\sqrt{c_{\mathbf{\sigma}}}}\sum_{\mathbf{b\in}\sigma}\operatorname{e}%
^{i\varphi_{\mathbf{b}}}\left\vert \mathbf{b}\right\rangle _{B}\left\vert
s\left(  \mathbf{b}\right)  \right\rangle _{A}\overset{_{M_{A}}}{\rightarrow
}\left\vert \mathbf{b}_{c}\right\rangle _{B}\left\vert s\left(  \mathbf{b}%
_{c}\right)  \right\rangle _{A}%
\]
We should advance this projection at the time of the initial measurement
$t_{1}$. By this we mean propagating the two end states of it backward in time
by the inverse of the unitary transformation that precedes it\footnote{The
notion of \textit{advancing} a state is taken from Wheeler-Feynman's absorber
theory $\left[  13\right]  $ and Cramer's transactional iterpretation of
quantum mechanics $\left[  14\right]  $.}, here by $U_{A}^{\dag}$. The
projection becomes:%
\[
\frac{1}{\sqrt{c_{\mathbf{\sigma}}}}\sum_{\mathbf{b\in}\sigma}\operatorname{e}%
^{i\varphi_{\mathbf{b}}}\left\vert \mathbf{b}\right\rangle _{B}\left\vert
\mathbf{0}\right\rangle _{A}\rightarrow\left\vert \mathbf{b}_{c}\right\rangle
_{B}\left\vert \mathbf{0}\right\rangle _{A},
\]
We can see that the projection that would have been performed by the initial
measurement is now performed back in time by the final measurement.

In $\left[  8,9\right]  $, we represented the subject ambiguity by generically
sharing the selection of the pair $\mathbf{b}_{c}$ and $s\left(
\mathbf{b}_{c}\right)  $ between the initial and final measurements. To this
end we ascribed the selection of a generic $\mathcal{R}$-th part of the
information that specifies the solution to the final measurement, that of the
complementary part to the initial measurement (in a quantum superposition of
all the possible ways of doing it as clarified below). By comparing the
consequent explanation of the speedup with a sample of optimal quantum
algorithms, it turned out that the sharing had to be even.

In the present work we go the other way around. We assume to start with that
the sharing is even. The reason is fundamental in character: any uneven
sharing would not be we invariant under time-reversal; it would introduce a
preferred direction of time, apparently with no reason in the present
completely reversible context (keeping in mind that there is a unitary
transformation between the initial and final measurement outcomes).

We note that this assumption is in line with (i) time-symmetric quantum
mechanics $\left[  14\div19\right]  $, which also excludes any preferred
direction of time, and (ii) the logical $\left[  20\right]  $ and physical
$\left[  21\div23\right]  $ reversibility of the computation process --
physical reversibility implies indifference to time-reversal.\ 

Evenly sharing the selection of the solution between the initial and final
measurements is inconsequential in the representation of the quantum algorithm
to Bob and any external observer. It would just say that part of the random
outcome of the initial measurement has been selected back in time by the final
measurement -- in fact an unverifiable and inconsequential thing.

In the representation to Alice things go differently. The former input state
to her (at time $t_{1+}$), of complete ignorance of the problem setting and
the corresponding solution, is projected on a state of lower entropy where she
knows one of the possible halves of the information that specifies them. The
projection has the form:%
\[
\frac{1}{\sqrt{c_{\mathbf{\sigma}}}}%
{\textstyle\sum\limits_{\mathbf{b\in}\sigma}}
\operatorname{e}^{i\varphi_{\mathbf{b}}}\left\vert \mathbf{b}\right\rangle
_{B}\left\vert \mathbf{0}\right\rangle _{A}\rightarrow\frac{1}{\sqrt
{c_{\mathbf{\sigma}^{\prime}}}}%
{\textstyle\sum\limits_{\mathbf{b\in}\sigma^{\prime}}}
\operatorname{e}^{i\varphi_{\mathbf{b}}}\left\vert \mathbf{b}\right\rangle
_{B}\left\vert \mathbf{0}\right\rangle _{A},
\]
where $\sigma^{\prime}$ is a suitable subset of $\sigma$.

Correspondingly, an optimal quantum algorithm should be seen as a sum over
classical histories in each of which Alice, the problem solver, knows in
advance one of the possible halves of the information about the solution she
will read in the future and performs the function evaluations still needed to
reach it.

We interpret Alice's advanced knowledge from a metric standpoint. We assume
that it gauges the distance, in number of function evaluations along the
classical histories, between the initial state of the quantum algorithm and
the solution. It would also be the distance covered by an optimal quantum
algorithm. Of course a non-optimal quantum algorithm would take a longer
route. As a matter of fact, all the oracle problems examined in this paper can
be solved quantumly with any number of function evaluations above that of the
optimal quantum algorithm.

Grover algorithm is optimal for any number of drawers $\left[  24\div
26\right]  $. The present retrocausal model exactly explains the speedup of
its four drawer instance, which requires just one function evaluation. Alice
knows in advance that the ball is in one of $2$ drawers. Opening either drawer
allows her to locate it. When greater than one, the number of function
evaluations required to find the solution along the classical histories is not
univocally defined; it depends on the search criteria. However, for any
reasonable criteria, the present retrocausal model gives the order of
magnitude of the number of function evaluations required by Grover algorithm.

All the other optimal quantum algorithms examined in this paper are solved
with just one function evaluation -- their speedup is exactly explained by the
present model.

Conversely, given an oracle problem, the model yields the order of magnitude
of the number of function \ evaluations needed to solve it in an optimal
quantum way. If this held for any oracle problem, as reasonable, it would
solve the well known open problem of quantum query complexity.

In section 3 we develop our explanation of the speedup on Grover algorithm. In
section 4 we generalize it to quantum oracle computing. \ In Section 5 we
apply the generalization to the algorithms of Deutsch\&Jozsa, Simon, and the
Abelian hidden subgroup. In sections 6 and 7, we discuss the present theory of
the speedup and provide the conclusions.

\section{Grover algorithm}

We show how the retrocausal explanation of the speedup comes from completing
the physical representation of Grover algorithm. This algorithm is best suited
to develop the present explanation, which is quantitative in character.
Besides being optimal, it involves a number of function evaluations growing
with problem size -- most of the quantum algorithms discovered so far require
just one function evaluation. We start with the four drawer instance of the algorithm.

\subsection{Time symmetric and relational representations}

Let the four drawers be numbered $00,01,10,11$ and $\mathbf{b}$ be the number
of the drawer with the ball. Checking whether the ball is in drawer
$\mathbf{a}$ amounts to computing the Kronecker function $f_{\mathbf{b}%
}\left(  \mathbf{a}\right)  \equiv\delta\left(  \mathbf{b},\mathbf{a}\right)
$, which is $1$ if $\mathbf{b}=\mathbf{a}$ and $0$ otherwise.

In the usual Grover algorithm, the number of the drawer\ that Alice wants to
open $\mathbf{a}$ (the argument of function evaluation) is contained in a
register $A$ of basis vectors $\left\vert 00\right\rangle _{A}$, $\left\vert
01\right\rangle _{A}$, $\left\vert 10\right\rangle _{A}$, $\left\vert
11\right\rangle _{A}$. This register, under the control of Alice, will
eventually contain the solution of the problem. A register $V$ (as value), of
basis vectors $\left\vert 0\right\rangle _{V},\left\vert 1\right\rangle _{V}$,
is meant to contain the result of function evaluation, modulo $2$ added to its
former content for logical reversibility. By the way, one could do without
this register: transformations remain unitary also without it -- but they are
simpler to describe with it.

As anticipated, the value of $\mathbf{b}$\ is not represented physically. Let
us assume it is $\mathbf{b}=01$. The usual Grover algorithm is limited to the
unitary transformation of the input state:%
\[
\left\vert \gamma\right\rangle =\frac{1}{\sqrt{2}}\left\vert 00\right\rangle
_{A}\left(  \left\vert 0\right\rangle _{V}-\left\vert 1\right\rangle
_{V}\right)
\]
into the output state%
\[
\Im_{A}U_{f}H_{A}\left\vert \gamma\right\rangle =\frac{1}{\sqrt{2}}\left\vert
01\right\rangle _{A}\left(  \left\vert 0\right\rangle _{V}-\left\vert
1\right\rangle _{V}\right)  ,
\]
where register $A$\ contains the solution of the problem -- namely the number
of the drawer with the ball $01$. $H_{A}$\ is the Hadamard transform on
register $A$. It transforms $\left\vert 00\right\rangle _{A}$\ into $\frac
{1}{2}\left(  \left\vert 00\right\rangle _{A}+\left\vert 01\right\rangle
_{A}+\left\vert 10\right\rangle _{A}+\left\vert 11\right\rangle _{A}\right)
$. $U_{f}$\ \ is function evaluation, thus performed in \textit{quantum
parallelism} $\left[  2\right]  $\ for all the possible values of $\mathbf{a}$
\footnote{The fact that Alice opens a single drawer for a quantum
superposition of all the possible drawer numbers is of course the key to
achieve the speedup, but it does not provide any quantitative explanation of
it in general.}. It leaves the state of register $V$, $\frac{1}{\sqrt{2}%
}\left(  \left\vert 0\right\rangle _{V}-\left\vert 1\right\rangle _{V}\right)
$, unaltered when $\mathbf{a}\neq01$ and thus $\delta=0$; it changes it into
$-~\frac{1}{\sqrt{2}}\left(  \left\vert 0\right\rangle _{V}-\left\vert
1\right\rangle _{V}\right)  $ when $\mathbf{a}=01$ and $\delta=1$\ (modulo $2$
addition of $1$\ changes $\left\vert 0\right\rangle _{V}$ into $\left\vert
1\right\rangle _{V}$ and vice-versa). The transformation $\Im_{A}$, applying
to register $A$,\ is the so called \textit{inversion about the mean }$\left[
1\right]  $: a rotation of the basis of $A$ makes the information acquired
with function evaluation accessible to measurement. We do not need to go into
further detail: all we need to know of the quantum algorithm is already there.

Eventually Alice acquires the solution by measuring the content of $A$, namely
the observable $\hat{A}$ of eigenstates the basis vectors of register $A$\ and
eigenvalues (correspondingly) $00$, $01$, $10$, $11$.

Now we extend the representation of the quantum algorithm to the process of
choosing the number of the drawer with the ball $\mathbf{b}$. We need to add a
possibly imaginary register $B$ that contains $\mathbf{b}$. This register,
under the control of Bob, has basis vectors $\left\vert 00\right\rangle _{B}$,
$\left\vert 01\right\rangle _{B}$, $\left\vert 10\right\rangle _{B}$,
$\left\vert 11\right\rangle _{B}$. We assume that the initial state of
register $B$ is a mixture of all the possible drawer numbers. To keep the
usual ket vector representation of quantum algorithms, we represent it as a
\textit{dephased} quantum state superposition:%
\[
\left\vert \psi\right\rangle _{B}=\frac{1}{2}\left(  \operatorname{e}%
^{i\varphi_{0}}\left\vert 00\right\rangle _{B}+\operatorname{e}^{i\varphi_{1}%
}\left\vert 01\right\rangle _{B}+\operatorname{e}^{i\varphi_{2}}\left\vert
10\right\rangle _{B}+\operatorname{e}^{i\varphi_{3}}\left\vert 11\right\rangle
_{B}\right)  .
\]
The $\varphi_{i}$ are independent completely random phases. The use we make of
the \textit{random phase representation} $\left[  10\right]  $ of a maximally
mixed state is trivial. Because of the character of the unitary
transformations involved (see below), we can see $\left\vert \psi\right\rangle
_{B}$ as it were a pure quantum state, like the $\varphi_{i}$\ were fixed
phases. Only when we need to compute its von Neumann entropy we need to
remember their random character. The von Neumann entropy of $\left\vert
\psi\right\rangle _{B}$ is $2$ bit. By the way, the usual density operator,
namely $\frac{1}{4}\left(  \left\vert 00\right\rangle _{B}\left\langle
00\right\vert _{B}+\left\vert 01\right\rangle _{B}\left\langle 01\right\vert
_{B}+...\right)  $, is the average over all the $\varphi_{i}$\ of $\left\vert
\psi\right\rangle _{B}\left\langle \psi\right\vert _{B}$.

The initial state of the three registers, at time $t_{0}$, is then:%
\begin{equation}
\left\vert \psi\right\rangle _{\mathcal{\Im}}=\frac{1}{2\sqrt{2}}\left(
\operatorname{e}^{i\varphi_{0}}\left\vert 00\right\rangle _{B}%
+\operatorname{e}^{i\varphi_{1}}\left\vert 01\right\rangle _{B}%
+\operatorname{e}^{i\varphi_{2}}\left\vert 10\right\rangle _{B}%
+\operatorname{e}^{i\varphi_{3}}\left\vert 11\right\rangle _{B}\right)
\left\vert 00\right\rangle _{A}\left(  \left\vert 0\right\rangle
_{V}-\left\vert 1\right\rangle _{V}\right)  . \label{ini}%
\end{equation}

Let the observable $\hat{B}$, of eigenstates the basis vectors of register $B$
and eigenvalues (correspondingly) $00$, $01$, $10$, and $11$, be the content
of register $B$ (by the way, $\hat{A}$ and $\hat{B}$\ commute).\ At time
$t_{0}$\ Bob measures $\hat{B}$\ selecting a drawer number at random, say
$10$. The state after measurement at time $t_{0+}$ is thus:%
\begin{equation}
\left\vert \psi\right\rangle =\frac{1}{\sqrt{2}}\left\vert 10\right\rangle
_{B}\left\vert 00\right\rangle _{A}\left(  \left\vert 0\right\rangle
_{V}-\left\vert 1\right\rangle _{V}\right)  . \label{am}%
\end{equation}
Then, by the unitary transformation $U_{B}$, he sends it into the state that
encodes the desired problem setting, say $01$. At time $t_{1}$, after $U_{B}$,
the state is thus:
\begin{equation}
U_{B}\left\vert \psi\right\rangle =\frac{1}{\sqrt{2}}\left\vert
01\right\rangle _{B}\left\vert 00\right\rangle _{A}\left(  \left\vert
0\right\rangle _{V}-\left\vert 1\right\rangle _{V}\right)  . \label{se}%
\end{equation}
$U_{B}$ can be for example:%

\[
U_{B}\equiv\left\vert 11\right\rangle \left\langle 00\right\vert
_{B}+\left\vert 10\right\rangle \left\langle 01\right\vert _{B}+\left\vert
01\right\rangle \left\langle 10\right\vert _{B}+\left\vert 00\right\rangle
\left\langle 11\right\vert _{B},
\]
it is simpler to think that it changes zeros into ones and ones into zeros.

State (\ref{se}) is the input state of the quantum algorithm -- with the ball
hidden by Bob in drawer $01$. Alice unitarily sends it into the output state:%
\begin{equation}
\Im_{A}U_{f}H_{A}U_{B}\left\vert \psi\right\rangle =\frac{1}{\sqrt{2}%
}\left\vert 01\right\rangle _{B}\left\vert 01\right\rangle _{A}\left(
\left\vert 0\right\rangle _{V}-\left\vert 1\right\rangle _{V}\right)  ,
\label{fi}%
\end{equation}
reached at time $t_{2}$. Note that the unitary part of Alice's action does not
change the content of register $B$, namely the problem setting chosen by Bob.
In fact $B$ is the \textit{control register} of the function evaluation
transformation $U_{f}$: each basis vector of $B$ (each value of $\mathbf{b}%
$)\ naturally affects the result of the computation of $f_{\mathbf{b}}\left(
\mathbf{a}\right)  $ while remaining unaltered throughout it. The other
unitary transformations of Alice's action, $H_{A}$ and $\Im_{A}$, do not apply
to $B$.

Eventually Alice acquires the solution of the problem by measuring $\hat{A}$.
Note that this measurement leaves the quantum state unaltered; there is thus a
unitary transformation between the initial and final measurement outcomes.

We can see that the representation of the quantum algorithm, now extended to
the process of choosing the problem setting, is still physically incomplete.
It is with respect to Bob, the problem setter, and any external observer, it
cannot be with respect to Alice, the problem solver. State (\ref{se}), with
register $B$\ in the sharp state $\left\vert 01\right\rangle _{B}$, would tell
her that the number of the drawer with the ball is $01$ -- it would tell her
the solution of the problem before she opens any drawer.

As well known, to Alice the value of $\mathbf{b}$ must be hidden inside the
black box that computes $\delta\left(  \mathbf{b},\mathbf{a}\right)  $. To
represent this concealment physically, we resort to the relational
interpretation of quantum mechanics $\left[  11,12\right]  $.

To Alice, we postpone the projection of the quantum state due to the initial
measurement until the end of the unitary part of her problem-solving action;
we should correspondingly \textit{retard}\ -- i. e. propagate forward in time
$\left[  13,14\right]  $ -- the two end states of the projection by the
unitary part in question.

As a consequence, the input state to Alice, at time $t_{1}$, remains the
initial state (\ref{ini}); in fact $U_{B}$, being a unitary transformation
applying to register $B$,\ leaves the maximally mixed state of this register
unaltered. The $2$ bit entropy of the state of register $B$ in the input state
of the quantum algorithm to Alice represents her complete ignorance of the
problem setting.

The unitary part of Alice's action sends this input state into the output
state:%
\begin{equation}
\Im_{A}U_{f}H_{A}U_{B}\left\vert \psi\right\rangle _{\mathcal{\Im}}=\frac
{1}{2\sqrt{2}}\left(
\begin{array}
[c]{c}%
\operatorname{e}^{i\varphi_{0}}\left\vert 00\right\rangle _{B}\left\vert
00\right\rangle _{A}+\operatorname{e}^{i\varphi_{1}}\left\vert 01\right\rangle
_{B}\left\vert 01\right\rangle _{A}\\
+\operatorname{e}^{i\varphi_{2}}\left\vert 10\right\rangle _{B}\left\vert
10\right\rangle _{A}+\operatorname{e}^{i\varphi_{3}}\left\vert 11\right\rangle
_{B}\left\vert 11\right\rangle _{A}%
\end{array}
\right)  \left(  \left\vert 0\right\rangle _{V}-\left\vert 1\right\rangle
_{V}\right)  . \label{out}%
\end{equation}
\ We can see that each possible problem setting is multiplied by the
corresponding solution. The final measurement of the content of register $A$
projects the output state (\ref{out}) on state (\ref{fi}), which is thus
common to the representation to Bob and any external observer and to the
representation to Alice. The measurement outcome is unpredictable to Alice as
usual -- not to Bob and any external observer who already know that the number
of the drawer with the ball is $\mathbf{b}=01$.

By the way, we note that the projection of the quantum state due to the
initial measurement, postponed until the end of the quantum algorithm, in the
present case of Grover algorithm coincides with that due to the final measurement.

\subsection{Sharing the selection of the random outcome of the initial
measurement between the initial and final measurements}

We provide two ways of sharing the selection of the random outcome of the
initial measurement between the initial and final measurements: the synthetic
way has a more evident physical meaning, the analytical way is more general
and can be applied also to quantum oracle computing.

\subsubsection{Synthetic method}

We reduce the complete measurements, of $\hat{B}$ and $\hat{A}$, to partial
measurements, of $\hat{B}_{i}$ and $\hat{A}_{j}$, such that they: (i) together
select without redundancies the random outcome of the initial measurement and
(ii) evenly contribute to the selection of the solution -- evenly reduce the
entropy of the reduced density operator of register $A$ in the output state.
Note that we are applying Occam's razor; we give up the condition that
everything is selected by the measurement performed first, not the economic
condition that there are no redundant selections.

For example, one way of sharing compatible with (i) and (ii) is assuming that
the initial measurement of $\hat{B}$ reduces to that of $\hat{B}_{0}$ (the
content of the left cell of register $B$) and the final measurement of
$\hat{A}$ to that of $\hat{A}_{1}$ (the content of the right cell of register
$A$). Note that the outcomes of the complete measurements should be left
unaltered, we should only share their selections between the two partial measurements.

We see how things go in the two representations of the quantum algorithm,
starting with that to Bob and any external observer. Here the measurement of
$\hat{B}_{0}$ at time $t_{0}$, selecting the left digit of the number
contained in register $B$, must select the $1$\ of the outcome of the initial
measurement $\mathbf{b}=10$. This\ projects the initial state (\ref{ini}) on:%
\begin{equation}
\left\vert \xi\right\rangle =\frac{1}{2}\left(  \operatorname{e}^{i\varphi
_{2}}\left\vert 10\right\rangle _{B}+\operatorname{e}^{i\varphi_{3}}\left\vert
11\right\rangle _{B}\right)  \left\vert 00\right\rangle _{A}\left(  \left\vert
0\right\rangle _{V}-\left\vert 1\right\rangle _{V}\right)  . \label{inbs}%
\end{equation}

At time $t_{2}$,\ state (\ref{inbs}) has evolved into:%
\begin{equation}
\Im_{A}U_{f}H_{A}U_{B}\left\vert \xi\right\rangle =\frac{1}{2}\left(
\operatorname{e}^{i\varphi_{2}}\left\vert 01\right\rangle _{B}\left\vert
01\right\rangle _{A}+\operatorname{e}^{i\varphi_{3}}\left\vert 00\right\rangle
_{B}\left\vert 00\right\rangle _{A}\right)  \left(  \left\vert 0\right\rangle
_{V}-\left\vert 1\right\rangle _{V}\right)  . \label{outb}%
\end{equation}
Then the measurement of $\hat{A}_{1}$, selecting the right digit of the number
of the drawer with the ball $01$ contained in register $A$, projects state
(\ref{outb}) on the original output state (\ref{fi}). Advancing the two ends
of this projection by the inverse of $\Im_{A}U_{f}H_{A}U_{B}$ projects state
(\ref{inbs}) on the original state after the initial measurement (\ref{am}).

Summing up, the two partial measurements rebuild the selection of the random
outcome of the initial measurement and that of the final measurement while
leaving the original quantum algorithm to Bob and any external observer
unaltered. Moreover, the measurement of $\hat{A}_{1}$ selects one of the two
bits that specify the solution, that of $\hat{B}_{0}$ -- retarded at the time
of the final measurement --\ the other bit. Therefore the selection of the
solution evenly shares between the two partial measurements as required.

In the quantum algorithm with respect to Bob and any external observer, evenly
sharing all selections between the initial and final measurements only says
that the right digit of the random outcome of the initial measurement has been
selected back in time by the final measurement, an unverifiable thing.
Retrocausality is inconsequential in this representation, which is the usual
one up to the representation of Bob's choice.

Things change dramatically in the representation with respect to Alice.

Here the measurement of $\hat{B}_{0}$ at time $t_{0}$\ selects the $1$\ of the
random outcome of the initial measurement\ $10$ without altering the original
quantum algorithm; in fact the projection of the quantum state associated with
it is postponed until the end of the algorithm. We can go directly to the
output state to Alice (\ref{out}), at time $t_{2}$, when the measurement of
$\hat{A}_{1}$ acquires the right digit of $\mathbf{a}=01$. This projects state
(\ref{out}) on:%

\begin{equation}
\left\vert \chi\right\rangle =\frac{1}{2}\left(  \operatorname{e}%
^{i\varphi_{1}}\left\vert 01\right\rangle _{B}\left\vert 01\right\rangle
_{A}+\operatorname{e}^{i\varphi_{3}}\left\vert 11\right\rangle _{B}\left\vert
11\right\rangle _{A}\right)  \left(  \left\vert 0\right\rangle _{V}-\left\vert
1\right\rangle _{V}\right)  . \label{altro}%
\end{equation}

Also now we should propagate this projection backward in time until it
selects, at time $t_{0+}$, the right digit of the outcome of the initial
measurement, namely the $0$ of $10$ -- the $1$ was selected by the measurement
of $\hat{B}_{0}$.

What is interesting to present ends is the value of this backward propagation
at time $t_{1}$, immediately after $U_{B}$ and before the unitary part of
Alice's action $\Im_{A}U_{f}H_{A}$. We should advance the two ends of the
projection of the output state (\ref{out}) on state (\ref{altro})\ by the
inverse of $\Im_{A}U_{f}H_{A}$. The result is the projection of the input
state (\ref{ini})\ on the state:%
\begin{equation}
H_{A}^{\dag}U_{f}^{\dag}\Im_{A}^{\dag}\left\vert \chi\right\rangle =\frac
{1}{2}\left(  \operatorname{e}^{i\varphi_{1}}\left\vert 01\right\rangle
_{B}+\operatorname{e}^{i\varphi_{3}}\left\vert 11\right\rangle _{B}\right)
\left\vert 00\right\rangle _{A}\left(  \left\vert 0\right\rangle
_{V}-\left\vert 1\right\rangle _{V}\right)  . \label{adv}%
\end{equation}

This is an outstanding consequence. State (\ref{adv}), the input state to
Alice under the assumption that the selection of the random outcome of Bob's
measurement evenly shares between the initial and final measurements,\ tells
her, before she performs any function evaluation, that the number of the
drawer with the ball is either $\mathbf{b}=01$ or $\mathbf{b}=11$. We will say
that Alice \textit{knows in advance} that $\mathbf{b}\in\left\{
01,11\right\}  $. We take this as a metric notion: advanced knowledge of half
of the problem setting, or of the corresponding half of the solution, gauges
the distance (in number of function evaluations along the classical histories)
of the input state from the solution -- we will further discuss this point in
sections 3.3 and 6.3.1.

\subsubsection{Analytical method}

Throughout this section, we put ourselves in the representation of the quantum
algorithm with respect to Alice.

The analytical method is a way of calculating Alice's advanced knowledge that
avoids the necessity of knowing the unitary part of Alice's action. It hinges
on the following points.

a) Since in the output state the content of register $A$ is a bijective
function of that of register $B$, a partial measurement of the content of $A$
can always be represented as a partial measurement of the content of $B$. We
can thus replace the measurement of the generic $\hat{A}_{j}$ in the output
state by that of the generic $\hat{B}_{j}$.

b) The reduced density operator of register $B$ and any projection thereof
remain unaltered throughout the unitary part of Alice's action, as we will see
in a moment.

In the random phase representation, the reduced density operator of register
$B$ is the trace over registers $A$ and $V$ of the overall state of the three
registers. In the input state of Alice's problem solving action (\ref{ini}),
it has evidently the form:%
\begin{equation}
\left\vert \psi\right\rangle _{B\ }=\frac{1}{2}\left(  \operatorname{e}%
^{i\varphi_{0}}\left\vert 00\right\rangle _{B}+\operatorname{e}^{i\varphi_{1}%
}\left\vert 01\right\rangle _{B}+\operatorname{e}^{i\varphi_{2}}\left\vert
10\right\rangle _{B}+\operatorname{e}^{i\varphi_{3}}\left\vert 11\right\rangle
_{B}\right)  . \label{psi}%
\end{equation}
This form remains unaltered throughout the unitary part of Alice's action --
namely until state (\ref{out}). This is because each basis vector of $B$ --
thus also any superposition thereof -- remains unaltered through it (see the
comment to equation \ref{fi}).

Measuring $\hat{B}_{1}$ -- now replacing $\hat{A}_{1}$ -- in the output state
(\ref{out}) selects the right digit of $01$, projecting $\left\vert
\psi\right\rangle _{B}$ on
\begin{equation}
\frac{1}{\sqrt{2}}\left(  \operatorname{e}^{i\varphi_{1}}\left\vert
01\right\rangle _{B}+\operatorname{e}^{i\varphi_{3}}\left\vert 11\right\rangle
_{B}\right)  . \label{ak}%
\end{equation}
Advancing this projection at time $t_{1}$, the time of the input state, by
$H_{A}^{\dag}U_{f}^{\dag}\Im_{A}^{\dag}$, leaves it unaltered.

Only the interpretation of the projection changes. Advanced at time $t_{1}$,
the projection of (\ref{psi}) on (\ref{ak}) becomes the projection of the
state of register $B$ in the input state of the quantum algorithm (\ref{ini}),
of complete ignorance of the problem setting, on the state of lower entropy
(\ref{ak}) that represents Alice's advanced knowledge.

Note that the projection in question can be obtained more simply by thinking
of measuring $\hat{B}_{1}$ in the input state (\ref{ini}). In fact any
measurement on the content of register $B$ can be seen as a projection of
$\left\vert \psi\right\rangle _{B}$\ and as such, for the purpose of the
present calculation, can be moved at will along the unitary part of Alice's action.

It is also convenient to move the measurement of $\hat{B}_{0}$ (the content of
the left cell of register $B$) from the initial state at time $t_{0}$ to the
input state at time $t_{1}$. To Alice, these two states are identical. It
suffices to ask that the measurement selects the left digit of the problem
setting chosen by Bob, namely of $01$, no more of the random outcome of the
initial measurement $10$.

We end up with the problem of splitting the measurement of $\hat{B}$ in the
input state to Alice into two partial measurements, of $\hat{B}_{i}$ and
$\hat{B}_{j}$, such that they (I) together select without redundancies the
problem setting chosen by Bob and (II) evenly contribute to the selection of
the solution -- evenly reduce the entropy of the reduced density operator of
register $A$\ in the output state.

At this point we have lost the memory of which was the measurement performed
in the output state. Therefore, once satisfied conditions (I) and (II), the
measurement of either $\hat{B}_{i}$ or $\hat{B}_{j}$ projects the reduced
density operator of register $B$ on an instance of Alice's advanced knowledge.
We say for short it projects $\sigma$\ -- the set of all the possible problem
settings -- on such an instance.

We will see in section 4\ that this way of calculating Alice's advanced
knowledge can be applied as it is to any oracle problem.

\subsection{Sum over classical histories}

We need to reconcile the notion of Alice's advanced knowledge of half of the
information that specifies the solution with the fact that such a half can be
taken in a plurality of ways; in other words we need to symmetrize the notion
for all the possible ways of taking half of the information. Moreover, we need
an operational interpretation of the notion.

We kill two birds with one stone by resorting to Feynman's path integral
formulation of quantum mechanics $\left[  27\right]  $. We see an optimal
quantum algorithm (its time-symmetric representation with respect to Alice) as
a sum over classical histories in each of which Alice knows in advance one of
the possible halves of the information about the solution and performs the
function evaluations needed to find the other half. An example of history is:
\[
\operatorname{e}^{i\varphi_{1}}\left\vert 01\right\rangle _{B}\left\vert
00\right\rangle _{A}\left\vert 0\right\rangle _{V}\overset{H_{A}}{\rightarrow
}\operatorname{e}^{i\varphi_{1}}\left\vert 01\right\rangle _{B}\left\vert
11\right\rangle _{A}\left\vert 0\right\rangle _{V}\overset{U_{f}}{\rightarrow
}\operatorname{e}^{i\varphi_{1}}\left\vert 01\right\rangle _{B}\left\vert
11\right\rangle _{A}\left\vert 0\right\rangle _{V}\overset{\Im_{A}%
}{\rightarrow}\operatorname{e}^{i\varphi_{1}}\left\vert 01\right\rangle
_{B}\left\vert 01\right\rangle _{A}\left\vert 0\right\rangle _{V}.
\]
The left-most state is one of the elements of the input\ state superposition
(\ref{ini}). The state after each arrow is one of the elements of the quantum
superposition generated by the unitary transformation of the state before the
arrow (with the exception of function evaluation which leaves the state
sharp); the transformation in question is specified above the arrow.

In the history we are dealing with, the problem setting is $\mathbf{b}=01$.
Register $B$\ is correspondingly in $\left\vert 01\right\rangle _{B}$
throughout Alice's action, which of course does not change the problem setting
chosen by Bob.

Alice performs function evaluation for $\mathbf{a}=11$. The content of
register $A$\ in the second and third state is correspondingly $\left\vert
11\right\rangle _{A}$: the basis vectors of both $B$ and $A$ always remain
unaltered through $U_{f}$. The state of register $V$\ also remains unaltered
here since the result of function evaluation is $\delta\left(  01,11\right)
=0$.

Alice's advanced knowledge must be that $\mathbf{b}$ belongs to $\left\{
01,11\right\}  $. In fact the subset of $\sigma$ that represents it to must
always comprise the problem setting chosen by Bob, namely $\mathbf{b}=01$; the
second element must be $\mathbf{b}=11$ given that Alice tries function
evaluation for $\mathbf{a}=11$. Since the state of register $V$ remains
unaltered through $U_{f}$, she knows that the result of function evaluation is
zero and thus that the ball must be in drawer $01$ -- this with just one
function evaluation.

Summing up, the sum over classical histories picture exactly explains the
speedup of the present four drawer instance of Grover algorithm. In the next
section we discuss the case of more than four drawers.

\subsection{Grover algorithm with $N>4$}

We go to the general case of $N=2^{n}$\ drawers -- $n$\ is the number of bits
that specify the drawer number. With Grover algorithm, the number of function
evaluations (drawer openings) required to find the solution is in general:
\[
k\left(  n\right)  =\frac{\pi}{4\arcsin2^{-n/2}}-\frac{1}{2}\simeq\frac{\pi
}{4}2^{n/2}.
\]

Let us call $%
\mathbb{N}
\left(  n\right)  $ the number of function evaluations foreseen by the present
retrocausal model of the speedup in the case of $2^{n}$ drawers. $%
\mathbb{N}
\left(  n\right)  $ has been defined as the number of function evaluations
needed to classically reach the solution given the advanced knowledge of one
of the possible halves of the information that specifies it. We should
distinguish between the two cases: $n=2$\ and $n>2$.

As we have seen, for $n=2$, we have $%
\mathbb{N}
\left(  2\right)  =$ $k\left(  2\right)  =1$. In words, the present
retrocausal model exactly explains the speedup of the four drawers instance of
Grover algorithm. The explanation is no more exact when $n>2$. While the
definition of Alice's advanced knowledge -- as half of the information that
specifies the solution -- remains unaltered,\ that of $%
\mathbb{N}
\left(  n\right)  $ becomes not univocal. It gets dependent on the criteria
adopted for searching the solution.

We provide a few examples. Given the advanced knowledge of $n/2$ of the bits
that specify the solution, we could define $%
\mathbb{N}
\left(  n\right)  $ as:

(i) The number of function evaluations required to have the a-priori certainty
of finding the solution through an exhaustive classical search (never using a
second time the same argument for function evaluation). In this case we would
have $%
\mathbb{N}
\left(  n\right)  =2^{n/2}-1$ function evaluations, against the $k\left(
n\right)  \simeq\frac{\pi}{4}2^{n/2}$ of the optimal Grover algorithm.

(ii) The average number of function evaluations required to classically reach
the solution under an exhaustive randomly ordered search. One can guess that
in this case $%
\mathbb{N}
\left(  n\right)  $ would be a bit smaller than $k\left(  n\right)  $.

(iii) The average number of function evaluations required to classically reach
the solution under a completely random search, etc.

Perhaps $%
\mathbb{N}
\left(  n\right)  $ could be defined more precisely by relating it to the
structure of the sum over classical histories. This should be for further
study. For the time being, we make reference to the exhaustive classical
search with a-priori certainty of finding the solution. This always gives the
order of magnitude of the number of function evaluations required by Grover algorithm.

By the way, all the above holds for Grover algorithm, which is optimal in
character. We should also note that there is always a quantum algorithm that
solves Grover's problem with any number of function evaluations provided it is
not smaller than the minimum number required by Grover algorithm. This is the
revision of Grover algorithm devised by Long\ $\left[  25,26\right]  $, which
can be tuned to solve Grover problem with any number of function evaluations
equal to or above the minimum required by Grover algorithm\footnote{The
algorithm devised by Long always yields the solution with absolute certainty,
Grover algorithm, with absolute certainty, only for $n=2$.}.

The two things go together as follows. We should keep in mind the assumption
that Alice's advanced knowledge gauges the distance, in number of function
evaluations along the classical histories, between the initial state of the
quantum algorithm and the solution. Of course it is also the distance covered
by the optimal quantum algorithm; non-optimal quantum algorithms can take any
longer route.

\section{Quantum oracle computing}

Until now, the retrocausal model of the speedup has been used to explain the
speedup of a known quantum algorithm. Now we show how to use it to calculate
the number of function evaluations needed to solve an oracle problem in an
optimal quantum way.

To start with, we assume that the solution is a bijective function of the
problem setting like in section 3.2. To calculate Alice's advanced knowledge
(hence the number of function evaluations), we need to find the pairs $\hat
{B}_{i}$ and $\hat{B}_{j}$ satisfying conditions (I) and (II) of section
3.2.2. We follow the analytical method of section 3.2.2, which dispenses us
from knowing the unitary part of Alice's action. We only need to know the
state of registers $B$ and $A$ in the input and output states of the
time-symmetric representation of the quantum algorithm with respect to Alice.
For any quantum algorithm that solves the problem, these have the form:%
\[
\left\vert \psi\right\rangle _{I\ }=\frac{1}{\sqrt{c_{\mathbf{\sigma}}}}%
\sum_{\mathbf{b\in\sigma}}e^{i\varphi_{\mathbf{b}}}\left\vert \mathbf{b}%
\right\rangle _{B}\left\vert \mathbf{0}\right\rangle _{A}\text{ and
}\left\vert \psi\right\rangle _{O\ }=\frac{1}{\sqrt{c_{\mathbf{\sigma}}}}%
\sum_{\mathbf{b\in\sigma}}e^{i\varphi_{\mathbf{b}}}\left\vert \mathbf{b}%
\right\rangle _{B}\left\vert \mathbf{s}\left(  \mathbf{b}\right)
\right\rangle _{A},
\]
where $\mathbf{b}$ and $\mathbf{s}\left(  \mathbf{b}\right)  $ are
respectively the setting and the solution of the problem; $\mathbf{b}$\ ranges
over the set of the problem settings $\sigma$ of cardinality
$c_{\mathbf{\sigma}}$. Note that $\left\vert \psi\right\rangle _{I}$ and
$\left\vert \psi\right\rangle _{O}$ are written uniquely on the basis of the
pairs $\mathbf{b}$ and $s\left(  \mathbf{b}\right)  $, namely of the oracle
problem; there is no need of knowing the unitary transformation in
between\footnote{By the way, there can always be such a unitary transformation
because the output $\mathbf{b},s\left(  \mathbf{b}\right)  $ conserves the
memory of the input $\mathbf{b}$.}.

Summing up, we can calculate Alice's advanced knowledge, and thus the number
of function evaluations required to solve the problem in an optimal quantum
way, solely on the basis of the problem itself.

We show that this method of calculation can be applied as it is also in the
case that the solution is a many to one function of the problem setting. To
satisfy condition (I) and (II), the measurements of $\hat{B}_{i}$ and $\hat
{B}_{j}$ must acquire two complementary halves of the information that
specifies the solution. However, they must also acquire information about the
problem setting that is not in the solution. It is the information that
identifies the problem setting chosen by Bob among the "many" settings that
correspond to the solution in question.

This raises the following objection. Since Alice (in each classical history)
knows in advance the information acquired by the measurement of either
$\hat{B}_{i}$ or $\hat{B}_{j}$, she also knows in advance information that is
not in the solution. Given that Alice knows in advance part of the information
she will acquire in the future, her knowing in advance information that is not
in the solution might seem in contradiction with the fact that she measures
only the solution. The way out is that, by measuring the solution, she also
triggers the projection of the quantum state due to the initial measurement of
$\hat{B}$, which cannot be retarded beyond the unitary part of her action.
Therefore, by measuring the solution, she necessarily acquires both the
solution and the problem setting chosen by Bob.

From now on, we call this method of calculating Alice's advanced knowledge and
thus the number of function evaluations required to solve an oracle problem in
an optimal quantum way \textit{the advanced knowledge rule}. In the following,
we apply it to a variety of oracle problems.

\section{Deutsch\&Jozsa, Simon, and the Abelian hidden subgroup algorithms}

We apply the advanced knowledge rule to compute the number of function
evaluations required to solve the oracle problems addressed by the other major
quantum algorithms, which are all optimal. We will always obtain the number
required by the real quantum algorithm.

\subsection{Deutsch\&Jozsa algorithm}

In Deutsch\&Jozsa problem, Bob chooses a function out of the set of all the
constant and \textit{balanced} functions (with the same number of zeroes and
ones) $f_{\mathbf{b}}\left(  \mathbf{a}\right)  :\left\{  0,1\right\}
^{n}\rightarrow\left\{  0,1\right\}  $. Array (\ref{dj}) gives the tables of
four of the eight functions for $n=2$:%
\begin{equation}%
\begin{tabular}
[c]{llllll}%
$\mathbf{a}$ & ${\small \,f}_{0000}\left(  \mathbf{a}\right)  $ &
${\small f}_{1111}\left(  \mathbf{a}\right)  $ & ${\small f}_{0011}\left(
\mathbf{a}\right)  $ & ${\small f}_{1100}\left(  \mathbf{a}\right)  $ & ...\\
00 & 0 & 1 & 0 & 1 & ...\\
01 & 0 & 1 & 0 & 1 & ...\\
10 & 0 & 1 & 1 & 0 & ...\\
11 & 0 & 1 & 1 & 0 & ...
\end{tabular}
\label{dj}%
\end{equation}
Note that we use the table of the function -- the sequence of function values
for increasing values of the argument -- as the suffix of the function. Alice
knows the set of functions but not Bob's choice and is to find whether the
function chosen by Bob is constant or balanced by computing $f_{\mathbf{b}%
}\left(  \mathbf{a}\right)  $ for appropriate values of $\mathbf{a}$.
Classically, this requires in the worst case a number of function evaluations
exponential in $n$. It requires just one function evaluation with the quantum
algorithm devised by Deutsch\&Jozsa $\left[  28\right]  $.

We use the advanced knowledge rule twice. First to explain the speedup of
Deutsch\&Jozsa algorithm. Then to compute the number of function evaluations
required to solve Deutsch\&Jozsa problem in an optimal quantum way; in this
latter case we must ignore Deutsch\&Jozsa algorithm.

We start with the first case. The time-symmetric representation of
Deutsch\&Jozsa algorithm with respect to Alice is:%
\[
\left\vert \psi\right\rangle _{I\ }=\frac{1}{4}\left(  \operatorname{e}%
^{i\varphi_{0}}\left\vert 0000\right\rangle _{B}+\operatorname{e}%
^{i\varphi_{1}}\left\vert 1111\right\rangle _{B}+\operatorname{e}%
^{i\varphi_{2}}\left\vert 0011\right\rangle _{B}+\operatorname{e}%
^{i\varphi_{3}}\left\vert 1100\right\rangle _{B}+...\right)  \left\vert
00\right\rangle _{A}\left(  \left\vert 0\right\rangle _{V}-\left\vert
1\right\rangle _{V}\right)  ,
\]%
\begin{align*}
H_{A}U_{f}H_{A}\left\vert \psi\right\rangle _{I\ }  &  =\frac{1}{4}\left[
\left(  \operatorname{e}^{i\varphi_{0}}\left\vert 0000\right\rangle
_{B}-\operatorname{e}^{i\varphi_{1}}\left\vert 1111\right\rangle _{B}\right)
\left\vert 00\right\rangle _{A}+\left(  \operatorname{e}^{i\varphi_{2}%
}\left\vert 0011\right\rangle _{B}-\operatorname{e}^{i\varphi_{3}}\left\vert
1100\right\rangle _{B}\right)  \left\vert 10\right\rangle _{A}+...\right] \\
&  \left(  \left\vert 0\right\rangle _{V}-\left\vert 1\right\rangle
_{V}\right)  .
\end{align*}
In general register $B$ is $2^{n}$ qubit, register $A$ is $n$\ qubit. $H_{A}$
is the Hadamard transform on register $A$, $U_{f}$\ is function evaluation. In
the output state, register $A$\ contains the pre-solution $s\left(
\mathbf{b}\right)  $: the function is constant if $s\left(  \mathbf{b}\right)
$ is all zeros, balanced otherwise. Measuring $\hat{A}$\ in the output\ state
yields $s\left(  \mathbf{b}\right)  $.

To compute Alice's advanced knowledge, we should split in all the possible
ways the initial measurement of $\hat{B}$ into two partial measurements, of
$\hat{B}_{i}$ and $\hat{B}_{j}$, satisfying conditions (I) and (II) of section 3.2.2.

Given the problem setting of a balanced function, there is only one pair of
partial measurements of the content of register $B$ compatible with these
conditions. With problem setting, say, $\mathbf{b}=0011$, $\hat{B}_{i}$ must
be the content of the left half of register $B$ and $\hat{B}_{j}$ that of the
right half. The measurement of $\hat{B}_{i}$ yields all zeros, that of
$\hat{B}_{j}$ all ones.

In fact, a partial measurement yielding both zeroes and ones would provide
enough information to identify the solution -- the fact that $f_{\mathbf{b}}$
is balanced. Then the cases are two. If the other partial measurement does not
contain both zeros and ones, it would not identify the solution; this would
violate the requirement that the two partial measurements evenly contribute to
the selection of the solution. If it did, the two partial measurements would
be redundant with one another.

Moreover, given that either partial measurement must yield all zeroes or all
ones, it must concern the content of half register. Otherwise either the
requirement of even contribution to the selection of the solution would be
violated or the problem setting would not be completely determined, as readily checked.

One can see that, with $\mathbf{b}$ $=0011$, the measurement of $\hat{B}_{i}$,
performed alone, projects $\sigma$ (the set of all the possible problem
settings) on the subset $\left\{  0011,0000\right\}  $, that of $\hat{B}_{j}$
on$\ \left\{  0011,1111\right\}  $. Either subset represents an instance of
Alice's advanced knowledge.

The case of the problem setting of a constant function is analogous. The only
difference is that there are more pairs of partial measurements that satisfy
conditions (I) and (II) -- see $\left[  7\right]  $.

There is a shortcut to finding the subsets of $\sigma$ that represent Alice's
advanced knowledge. Here the problem setting -- the bit string $\mathbf{b}$ --
is the table of the function chosen by Bob. For example $\mathbf{b}=0011$ is
the table $f_{\mathbf{b}}\left(  00\right)  =0,f_{\mathbf{b}}\left(
01\right)  =0,f_{\mathbf{b}}\left(  10\right)  =1,f_{\mathbf{b}}\left(
11\right)  =1$. We call \textit{good half table} any half table in which all
the values of the function are the same. One can see that good half tables are
in one-to-one correspondence with the subsets in question. For example, the
good half table $f_{\mathbf{b}}\left(  00\right)  =0,f_{\mathbf{b}}\left(
01\right)  =0$ corresponds to the subset $\left\{  0011,0000\right\}  $, is
the identical part of the two bit-strings in it. Thus, given a problem
setting, i. e. an entire table, either good half table, or identically the
corresponding subset of $\sigma$, is a possible instance of Alice's advanced knowledge.

Because of the structure of tables, given the advanced knowledge of a good
half table, the entire table and thus the solution can be identified by
performing just one function evaluation for any value of the argument
$\mathbf{a}$ outside the half table.

Summing up, the advanced knowledge rule explains the fact that Deutsch\&Jozsa
algorithm requires just one function evaluation -- it explains the algorithm's
exponential speedup. By the way, for the fact of requiring just one function
evaluation, this quantum algorithm is necessarily optimal.

One can see that the present analysis, like the notion of good half table,
holds unaltered for $n>2$.

We show how to apply the advanced knowledge rule to compute the number of
function evaluations required to solve Deutsch\&Jozsa problem in an optimal
quantum way, of course without knowing Deutsch\&Jozsa algorithm.
Now\ $s\left(  \mathbf{b}\right)  $\ is the solution of the original problem:
$0$\ if the function is constant and $1$\ if balanced. The input and output
states of the registers $B$ and $A$ of any quantum algorithm that solves the
problem must be respectively:%
\[
\left\vert \psi\right\rangle _{I\ }=\frac{1}{2\sqrt{2}}\left(
\operatorname{e}^{i\varphi_{0}}\left\vert 0000\right\rangle _{B}%
+\operatorname{e}^{i\varphi_{1}}\left\vert 1111\right\rangle _{B}%
+\operatorname{e}^{i\varphi_{2}}\left\vert 0011\right\rangle _{B}%
+\operatorname{e}^{i\varphi_{3}}\left\vert 1100\right\rangle _{B}+...\right)
\left\vert 0\right\rangle _{A}%
\]%
\[
\left\vert \psi\right\rangle _{O\ }=\frac{1}{2\sqrt{2}}\left[  \left(
\operatorname{e}^{i\varphi_{0}}\left\vert 0000\right\rangle _{B}%
-\operatorname{e}^{i\varphi_{1}}\left\vert 1111\right\rangle _{B}\right)
\left\vert 0\right\rangle _{A}+\left(  \operatorname{e}^{i\varphi_{2}%
}\left\vert 0011\right\rangle _{B}-\operatorname{e}^{i\varphi_{3}}\left\vert
1100\right\rangle _{B}\right)  \left\vert 1\right\rangle _{A}+...\right]  .
\]

Note that $\left\vert \psi\right\rangle _{I}$ and $\left\vert \psi
\right\rangle _{O}$ have been written uniquely on the basis of the oracle
problem, namely of the pairs $\mathbf{b}$ and $s\left(  \mathbf{b}\right)  $.

One can readily see that Alice's advanced knowledge, and thus the number of
function evaluations required to solve the problem in an optimal quantum way,
can be computed exactly as in the case of Deutsch\&Jozsa algorithm. All
results are the same.

\subsection{Simon and the Abelian hidden subgroup algorithms}

Simon problem consists in finding the "period" (up to bitwise modulo 2
addition) of a periodic function $f_{\mathbf{b}}\left(  \mathbf{a}\right)
:\left\{  0,1\right\}  ^{n}\rightarrow\left\{  0,1\right\}  ^{n-1}$ -- see
$\left[  7\right]  $ for details. Array (\ref{periodic}) gives the tables of
four of the six functions for $n=2$:%
\begin{equation}%
\begin{tabular}
[c]{llllll}%
$\mathbf{a}$ & ${\small f}_{0011}\left(  \mathbf{a}\right)  $ & ${\small f}%
_{1100}\left(  \mathbf{a}\right)  $ & ${\small f}_{0101}\left(  \mathbf{a}%
\right)  $ & ${\small f}_{1010}\left(  \mathbf{a}\right)  $ & ...\\
00 & 0 & 1 & 0 & 1 & ...\\
01 & 0 & 1 & 1 & 0 & ...\\
10 & 1 & 0 & 0 & 1 & ...\\
11 & 1 & 0 & 1 & 0 & ...
\end{tabular}
\label{periodic}%
\end{equation}

We note that each value of the function appears exactly twice in each table;
thus 50\% of the rows plus one always identify the period. Alice is to find
the period of the function by performing function evaluation\ for appropriate
values of $\mathbf{a}$.

In present knowledge, a classical algorithm requires a number of function
evaluations exponential in $n$. The quantum part of Simon algorithm $\left[
29\right]  $ solves with just one function evaluation the hard part of this
problem, which is finding a bit string \textit{orthogonal} to the period. See
$\left[  7\right]  $ for further detail.

We apply the advanced knowledge rule directly to the calculation of the number
of function evaluations required to solve Simon problem in an optimal quantum
way. We will see further below that this also explains the speedup of Simon algorithm.

The input and output states of the registers $B$ and $A$ of any quantum
algorithm that solves the problem must be respectively:%

\[
\left\vert \psi\right\rangle _{I\ }=\frac{1}{\sqrt{6}}\left(  \operatorname{e}%
^{i\varphi_{0}}\left\vert 0011\right\rangle _{B}+\operatorname{e}%
^{i\varphi_{1}}\left\vert 1100\right\rangle _{B}+\operatorname{e}%
^{i\varphi_{2}}\left\vert 0101\right\rangle _{B}+\operatorname{e}%
^{i\varphi_{3}}\left\vert 1010\right\rangle _{B}+...\right)  \left\vert
00\right\rangle _{A},
\]%
\[
\left\vert \psi\right\rangle _{O\ }=\frac{1}{\sqrt{6}}\left[  \left(
\operatorname{e}^{i\varphi_{0}}\left\vert 0011\right\rangle _{B}%
+\operatorname{e}^{i\varphi_{1}}\left\vert 1100\right\rangle _{B}\right)
\left\vert 01\right\rangle _{A}+\left(  \operatorname{e}^{i\varphi_{2}%
}\left\vert 0101\right\rangle _{B}+\operatorname{e}^{i\varphi_{3}}\left\vert
1010\right\rangle _{B}\right)  \left\vert 10\right\rangle _{A}+...\right]  .
\]
In the output state, the $2^{n}$ qubit register $B$ contains the problem
setting $\mathbf{b}$ and the $n$\ qubit register$\ A$ the corresponding
solution of the problem $s\left(  \mathbf{b}\right)  $, namely the period of
the function ${\small f}_{\mathbf{b}}$.

Here a good half table, which represents an instance of Alice's advanced
knowledge like in Deutsch\&Jozsa algorithm, is any half table where the values
of the function are all different from one another (so that the period cannot
be identified) -- see $\left[  7\right]  $. Since 50\% of the rows plus one
identify the period, this can always be found by performing just one function
evaluation for any value of the argument $\mathbf{a}$\ outside the good half table.

Summing up, the advanced knowledge rule says that Simon's problem can be
solved with just one function evaluation. This also explains the exponential
speedup of Simon algorithm. Of course, according to the advanced knowledge
rule, also finding a bit string orthogonal to the period requires just one
function evaluation (knowing the period amounts to know all the bit strings
orthogonal to it).

The present analysis, like the notion of good half table, holds unaltered for
$n>2$. It should also apply to the generalized Simon's problem and to the
Abelian hidden subgroup problem. In fact the corresponding algorithms are
essentially Simon algorithm. In the Abelian hidden subgroup problem, the set
of functions $f_{\mathbf{b}}:G\rightarrow W$ map a group $G$ to some finite
set $W$\ with the property that there exists some subgroup $S\leq G$ such that
for any $\mathbf{a},\mathbf{c}\in G$, $f_{\mathbf{b}}\left(  \mathbf{a}%
\right)  =f_{\mathbf{b}}\left(  \mathbf{c}\right)  $ if and only if
$\mathbf{a}+S=\mathbf{c}+S$. The problem is to find the hidden subgroup $S$ by
computing $f_{\mathbf{b}}\left(  \mathbf{a}\right)  $ for the appropriate
values of $\mathbf{a}$. Now, a large variety of problems solvable with a
quantum speedup can be re-formulated in terms of the Abelian hidden subgroup
problem. Among these we find: the seminal Deutsch problem, finding orders,
finding the period of a function (thus the problem solved by the quantum part
of Shor\ factorization algorithm), discrete logarithms in any group, hidden
linear functions, self shift equivalent polynomials, Abelian stabilizer
problem, graph automorphism problem $\left[  30\right]  $.

\section{Discussion}

As we have moved into uncharted waters, it is worth discussing at some length
the present retrocausal explanation of the speedup.\ \ \ \ \ 

\subsection{Positioning}

The present explanation of the speedup relies on three areas of research that
had remained separate until now: (i) time-symmetric quantum mechanics, (ii)
relational quantum mechanics, and (iii)\ quantum computation.

The notion that the complete description of the quantum computation process
must include a forward propagation from the initial measurement and a backward
propagation from the final one, central to the present explanation of the
speedup, has clearly been inspired by time-symmetric quantum mechanics
$\left[  13\div19\right]  $. It is a variation of its standard form in the
particular case that there is a unitary transformation between the initial and
final measurement outcomes and that the initial measurement is performed in a
maximally mixed state. We represent the indifference of ascribing the
selection of the random outcome of the initial measurement to the initial or
final measurement by sharing it evenly between the two. This particular
formalization of the time-symmetric picture has been inspired by the work of
Dolev and Elitzur on the non-sequential behavior of the wave function
highlighted by partial measurement $\left[  16\right]  $.

The role played by relational quantum mechanics $\left[  11,12\right]  $ is
equally essential. In the representation of the quantum algorithm with respect
to Bob (the problem setter) and any external observer, which is close to the
usual representation, retrocausality is without consequences. In the
representation with respect to Alice (the problem solver) it explains the speedup.

For what concerns the study of the speed up, the present approach is
orthogonal to the mainstream ones. As far as we know, no other approach
resorts to the notion of retrocausality.

There are various studies on the relationship between speedup and other
fundamental quantum features such as entanglement and discord $\left[
31\div34\right]  $. However, until now, these studies could not provide a
common explanation to the various speedups. Quoting from $\left[  32\right]
$: \textit{The speedup appears to always depend on the exact nature of the
problem while the reason for it varies from problem to problem}.

The present retrocausal interpretation of the speedup, instead, quantitatively
justifies both the quadratic and a variety of exponential speedups. Moreover,
given an oracle problem, the advanced knowledge rule foresees the order of
magnitude of the number of function evaluations required to solve it in an
optimal quantum way -- this at least in the very diverse cases examined.

For completeness, we mention the other main approaches to the study of the
speedup: (i) Quantum computer science. Its aim is to find where quantum
complexity classes, such as BQP and QMA, lie with respect to classical
complexity classes such as P, NP, PP, etc. -- see $\left[  35\right]  $ for an
example. (ii) Tree size complexity. A measure of the complexity of the
multiqubit state is shown to be related to the speedup of a variety of quantum
algorithms $\left[  36\right]  $. (iii) Contextually based arguments, which
address the relation between speedup and the contextual character of quantum
mechanics $\left[  37\right]  $. The present approach would be in competition
with these others in providing some estimate of the number of function
evaluations required to solve an oracle problem in an optimal quantum way. Its
promise, providing an order of magnitude estimate for any oracle problem,
would be unparalleled by the other approaches.

We should also mention a work of Morikoshi $\left[  38\right]  $ that might be
related to our own. It shows that Grover algorithm violates an information
theoretic temporal Bell inequality. The present notion that Alice knows in
advance -- in each classical history -- half of the information about the
solution she will read in the future, a form of temporal nonlocality, is
likely related to the violation in question. This should be for further study.

The present interpretation of the speedup is also in line with those
explanations of quantum nonlocality that resort to the notion of
retrocausality -- see $\left[  39,40\right]  $. Causality \textit{zigzagging}
back and forth between the measurements of two entangled observables is a
common feature.

\subsection{Grover's anticipation}

Interestingly, Grover anticipated the need for a simple explanation of the
speedup $\left[  41\right]  $. Quoting his words: \textit{It has been proved
that the quantum search algorithm cannot be improved at all, i.e. any quantum
mechanical algorithm will need at least} $\operatorname*{O}\left(  \sqrt
{N}\right)  $\textit{\ steps to carry out an exhaustive search of }%
$N$\textit{\ items [4] [5]. Why is it not possible to search in fewer than
}$\operatorname*{O}\left(  \sqrt{N}\right)  $\textit{\ steps? The arguments
used to prove this are very subtle and mathematical. What is lacking is a
simple and convincing two line argument that shows why one would expect this
to be the case. }

The present "two line argument" would be that the quantum search algorithm is
a sum over classical histories in each of which the problem solver knows in
advance one of the possible halves of the information that specifies the
solution she will read in the future and performs the function evaluations
still needed to reach it.

That the advanced knowledge of half of the information about the number of the
drawer with the ball explains the quadratic speedup of Grover algorithm is
almost tautological. The important thing is that this is the seed of a more
general notion. We have seen that knowing in advance half of the information
about the solution explains as well the major exponential speedups and seems
to answer a fundamental time-reversal symmetry.\qquad\qquad\qquad\qquad
\qquad\qquad\qquad\qquad\qquad\qquad\qquad\qquad\qquad\qquad\qquad\qquad
\qquad\qquad\qquad\qquad\qquad\qquad\qquad\qquad\qquad\qquad\qquad\qquad
\qquad\qquad\qquad\qquad\qquad\qquad\qquad\qquad\qquad\qquad\qquad\qquad
\qquad\qquad\qquad\qquad\qquad\qquad\qquad\qquad\qquad\qquad\qquad\qquad
\qquad\qquad\qquad\qquad\qquad\qquad\qquad

\subsection{Criticism of retrocausality}

We discuss the present interpretation of the speedup at the light of the
criticism typically moved to the use of retrocausality in physics.

\subsubsection{Bell's criticism}

Quoting from $\left[  42\right]  $: \textit{As Bell was well aware, the
dilemma} [of non locality] \textit{can be avoided if the properties of quantum
systems are allowed to depend on what happens to them in the future, as well
as in the past. Like most researchers interested in these issues, however,
Bell felt that the cure would be worse than the disease -- he thought that
this kind of \textquotedblleft retrocausality\textquotedblright\ would
conflict with free will, and with assumptions fundamental to the practice of
science. (He said that when he tried to think about retrocausality, he
\textquotedblleft lapsed into fatalism\textquotedblright).} See also $\left[
29,30\right]  $. \ 

We compare the present interpretation of the speedup with Bell's observations.

We have seen that retrocausality is without consequences in the quantum
algorithm with respect to Bob and any external observer. It instead affects
the quantum algorithm with respect to Alice. In the four drawers instance of
Grover algorithm, the projection induced by the final measurement of $\hat
{A}_{1}$ in state (\ref{out}), propagating backward in time, at time $t_{1}$
projects the input state to her (\ref{ini})\ on:%
\[
H_{A}^{\dag}U_{f}^{\dag}\Im_{A}^{\dag}\left\vert \chi\right\rangle =\frac
{1}{2}\left(  \operatorname{e}^{i\varphi_{1}}\left\vert 01\right\rangle
_{B}+\operatorname{e}^{i\varphi_{3}}\left\vert 11\right\rangle _{B}\right)
\left\vert 00\right\rangle _{A}\left(  \left\vert 0\right\rangle
_{V}-\left\vert 1\right\rangle _{V}\right)  ,
\]
At first sight, one may think that this projection, selecting the right digit
of the number of the drawer with the ball, restricts back in time Bob's
freedom of choice to choosing between $\mathbf{b}=01$ and $\mathbf{b}=11$.\ 

However, this would require that the projection in question is free -- random
-- and independent of Bob's choice. This is not the case. In spite of the fact
that the measurement of $\hat{A}_{1}$ is performed in the mixture of tensor
products (\ref{out}), the associated projection of the quantum state occurs
deterministically on the right digit of the number of the drawer with the ball
already chosen by Bob at time $t_{1}$. We should keep in mind that we are in
the quantum algorithm with respect to Alice. The outcome of the final
measurement is unpredictable to her, not to Bob and any external observer who
already know the number of the drawer with the ball.

Summing up, advancing the projection in question does not restrict Bob's
choice but only Alice's ignorance of it.

We face now the other Bell's observation, about the possible conflict of
retrocausality with assumptions fundamental to the practice of science. Let us
refer again to the four drawer instance of Grover algorithm. The fact that
Alice, in each classical history, knows in advance half of the information
about the number of the drawer with the ball she will read in the future is an
obvious candidate to conflict. Obvious questions are: (i) Alice is an abstract
entity, what does it mean that "she knows"? (ii) At time $t_{1}$\ Alice has
done nothing yet, what tells her this information? (iii) Is it information
sent back in time?

The answer to question (i) is that we are at a fundamental level where knowing
is doing $\left[  43\right]  $. Alice, the problem solver, "knows" from an
operational standpoint, as far as the solution can be reached with a
correspondingly reduced number of function evaluations. In other words we are
talking of the metric of quantum computation. Alice's advanced knowledge in
the input state of the quantum algorithm gauges, in number of function
evaluations, the distance of this state from the solution.

About question (ii), Alice is told by the backward propagation of the
projection induced by the final measurement of $\hat{A}_{1}$. On its way to
selecting, at time $t_{0+}$, part of the random outcome of the initial
measurement, at the intermediate time $t_{1}$ it projects the original input
state to Alice, of complete ignorance of the problem setting and consequently
the solution, on one of lower entropy where she knows part of them -- in the
operational way discussed above.

Question (iii) is whether Alice's advanced knowledge -- in each classical
history -- of part of the information about the solution she will read in the
future implies that information is sent back in time. The same question
applies to the related fact that part of the random outcome of the initial
measurement is selected back in time by the final measurement. Of course there
is information sent back in time along the classical histories, but our answer
is negative as long as it is not measurable.

At time $t_{1}$, immediately before the beginning of Alice's problem-solving
action, the information in question would be in register $B$. It would
correspond to a reduction of the entropy of the state of this register, in
fact to a reduction of Alice's ignorance of the problem setting.

By definition, the observer Alice cannot perform any measurement of the
content of register $B$ at time $t_{1}$. If she did, she would destroy the
physical context that originates her advanced knowledge.

Bob and any external observer instead do measure the content of register
$B$\ at time $t_{0}$ and could repeat the measurement at time $t_{1}$. At time
$t_{0+}$, they see a completely random measurement outcome and have no way of
saying whether part of it is selected back in time by the final measurement.
At time $t_{1}$, they would see the problem setting freely chosen by Bob. In
either case, no information coming from the future can be identified in the
measurement outcome.

We would like to add a common sense consideration. The idea that Alice, in
each classical history, knows in advance part of what she will read in the
future might anyway conflict with our sense of physical reality. Our advice
for the time being would be to stick to the quantum computation context, where
Alice's advanced knowledge has a precise meaning and admits an apparently
harmless metric interpretation -- anyway no longer harmful than the speedup itself.

\subsubsection{Redundancy of the retrocausal interpretation}

A natural question is whether the time-symmetric and relational
interpretations of quantum mechanics are necessary to derive the present
results. An upstream question is of course whether these interpretations are
necessary to quantum mechanics. Apropos of the latter question we cite the
following positions.

Rovelli drew an analogy between relational quantum mechanic and special
relativity; in both cases physical quantities must be related to the observer.
After noting the revolutionary impact of the famous 1905 Einstein's paper,
Rovelli\textit{ }$\left[  12\right]  $ writes: \textit{The formal content of
special relativity, however is coded into the Lorentz transformations, written
by Lorentz, not by Einstein, and before 1905. So, what was Einstein's
contribution? It was to understand the physical meaning of the Lorentz
transformations. (And more, but this is what is of interest here). We could
say -- admittedly in a provocative manner -- that Einstein's contribution to
special relativity has been the interpretation of the theory, not its
formalism: the formalism already existed. }

Elitzur expressed a similar position about the time symmetric interpretations
of quantum mechanics $\left[  44\right]  $: even if they were pure
interpretations, adding nothing to the formalism, they did and could allow to
see things that would be otherwise very difficult to see.

We believe that the retrocausal interpretation of the speedup lends itself to
similar considerations.

The number of function evaluations required to solve an oracle problem in an
optimal quantum way, presumably given in the order of magnitude by the
advanced knowledge rule, should also be implicit in the mathematics of unitary
transformations, as follows.

In the most general case, the transformation that represents the unitary part
of Alice's problem solving action is (in the appropriate Hilbert space) a
sequence of function evaluations, each preceded and followed by a suitable
unitary transformation.

In principle, these transformations could be seen as the unknowns of the
problem of finding the optimal quantum algorithm. They should have variable
matrix elements up to the unitarity of the transformation.

For a given number of function evaluations, we should find the values of the
matrix elements that maximize the probability of finding the solution in the
final measurement; then repeat the procedure each time with that number
increased by one; stop when the probability in question reaches one.

We would have obtained analytically the number of function evaluations
required by the optimal quantum algorithm. In present assumptions, the order
of magnitude of this number is given in a synthetic way by the advanced
knowledge rule.

However, the analytic way is likely impracticable. In this case the synthetic
one, based on the time-symmetric and relational interpretations of quantum
mechanics, could provide a stunning shortcut; it does in all the cases examined.

\section{Conclusions}

The principle of the present explanation of the speedup is simple. Let us use
again the simplifying assumption that the solution is a one to one function of
a problem setting that is directly the random outcome of the initial measurement.

The selection of the problem setting and the corresponding solution can be
performed indifferently by the initial measurement of the problem setting or
the final measurement of the solution. We share it evenly between the two --
any uneven sharing would introduce a preferred direction of time apparently
unjustified in the present fully reversible context.

In the representation of the quantum algorithm to Bob, the problem setter, and
any external observer, this says that half of the information that specifies
the random outcome of the initial measurement has been selected by the final
measurement, an inconsequential thing.

In that to Alice, the problem solver, it projects the input state of the
quantum algorithm to her, of complete ignorance of the problem setting and the
solution, on one of lower entropy where she knows one of the possible halves
of the information that specifies them. An optimal quantum algorithm turns out
to be a sum over classical histories in each of which Alice knows in advance
one of the possible halves of the information about the solution she will read
in the future and performs the function evaluations still needed to reach it.

We interpret Alice's advanced knowledge in a metric way. It would gauge the
distance, in number of function evaluations along the classical histories, of
the input state of the quantum algorithm from the solution.

Conversely, given an oracle problem, the number of function evaluations
required to solve it in an optimal quantum way is that of a classical
algorithm that knows in advance half of the information about the solution of
the problem.

Summing up, although just a physical interpretation of the mathematics of
quantum algorithms, the present explanation of the speedup has potentially
important practical consequences. Until now there was no fundamental
explanation of the speedup, no unification of the quadratic and exponential
speedups, no solution to the so called quantum query complexity problem. The
subject explanation provides a fundamental, quantitative justification of all
kinds of speedup and promises to solve the problem in question.

The present form of quantum retrocausality, which cannot be measured but can
explain the higher than classical efficiency of a quantum process, might be
interesting from the foundational standpoint.

\subsection{Acknowledgments}

I had useful discussions with David Ritz Finkelstein along the entire
development of the present theory of the speedup. Thanks for useful comments
are due to Elihau Cohen, Avshalom Elitzur, and Kenneth Wharton.

\subsection*{References}

\ \ \ \ \ $\left[  1\right]  $ Grover, L. K.: A fast quantum mechanical
algorithm for database search. Proc. 28th Annual ACM Symposium on the Theory
of Computing; ACM press New York 212-219 (1996)

$\left[  2\right]  $ Deutsch, D.: Quantum theory, the Church Turing principle
and the universal quantum computer. Proc. Roy. Soc. A 400, 97-117(1985)

$\left[  3\right]  $ Feynman, R. P.: Simulating Physics with Computers. Int.
J. Theor. Phys. VoL 21, Nos. 6/7,\ 467-488 (1982)

$\left[  4\right]  $ Castagnoli, G. and Finkelstein, D. R.: Theory of the
quantum speedup.\textit{\ }Proc. Roy. Soc. A 1799, 457, 1799-1807 (2001)

$\left[  5\right]  $\ Castagnoli, G.: The quantum correlation between the
selection of the problem and that of the solution sheds light on the mechanism
of the quantum speed up. Phys. Rev. A 82, 052334-052342 (2010)

$\left[  6\right]  $ Castagnoli, G.: Probing the mechanism of the quantum
speed-up by time-symmetric quantum mechanics. Proceedings of the 92nd\ Annual
Meeting of the Pacific Division of the American Association for the
Advancement of Science, Quantum Retrocausation II, Program organizer Daniel
Sheehan (2011)

$\left[  7\right]  $ Castagnoli, G.: Highlighting the mechanism of the quantum
speedup by time-symmetric and relational quantum mechanics. Found. Phys. Vol.
46, Issue 3. 360--381 (2016)

$\left[  8\right]  $ Castagnoli, G.: On the relation between quantum
computational speedup and retrocausality. Quanta Vol. 5, No 1, 34-52 (2016)

$\left[  9\right]  $ Castagnoli, G.: A retrocausal model of the quantum
computational speedup. Proceedings of the 92nd\ Annual Meeting of the Pacific
Division of the American Association for the Advancement of Science, Quantum
Retrocausation III, Program organizer Daniel Sheehan (2016)

$\left[  10\right]  $ Bohm, D. and Pines, D. A.: A Collective description of
electron interactions. Coulomb interactions in a degenerate electron gas.
Phys. Rev. 92, 626-636 (1953)

$\left[  11\right]  $\ Rovelli, C.: Relational Quantum Mechanics. Int. J.
Theor. Phys. 35,\textbf{\ }637-658 (1996)

$\left[  12\right]  $ Rovelli C.: Relational Quantum Mechanics (2011)\ http://xxx.lanl.gov/pdf/quant-ph/9609002v2

$\left[  13\right]  $ Wheeler, J. A. and Feynman, R. P.: Interaction with the
Absorber as the Mechanism of Radiation. Rev. Mod. Phys. 17, 157 (1945)

$\left[  14\right]  $ Cramer J. "The Transactional Interpretation of Quantum
Mechanics" Rev. Mod. Phys. 58, 647 (1986)

$\left[  15\right]  $\ Aharonov, Y., Bergman, P. G., and Lebowitz, J. L.: Time
Symmetry in the Quantum Process of Measurement. Phys. Rev. B 134,\textbf{\ }%
1410-1416 (1964)

$\left[  16\right]  $\ Dolev, S. and Eitzur, A. C.: Non-sequential behavior of
the wave function. arXiv:quant-ph/0102109 v1 (2001)

$\left[  17\right]  $ Aharonov, Y. and Vaidman, L.: The Two-State Vector
Formalism: An Updated Review. Lect. Notes Phys. 734, 399--447 (2008)

$\left[  18\right]  $ Aharonov, Y., Popescu, S., and\ Tollaksen, J.: A
time-symmetric formulation of quantum mechanics. Physics Today, November issue
27-32 (2010)

$\left[  19\right]  $ Aharonov, Y., Cohen, E., Grossman, D., and Elitzur, A.
C.: Can a Future Choice Affect a Past Measurement's Outcome? arXiv:1206.6224 (2012)

$\left[  20\right]  $ Fredkin, E., Toffoli, T.: Conservative Logic. Int. J.
Theor. Phys. 21, 219-253 (1982)

$\left[  21\right]  $ Landauer, R.: Irreversibility and heat generation in the
computing process. IBM Journal of Research and Development, 5 (3), 183--191, 1961

$\left[  22\right]  $ Finkelstein, D. R.: Space-time structure in high energy
interactions. In Gudehus T, Kaiser G, Perlmutter A editors, Fundamental
Interactions at High Energy. New York: Gordon \& Breach, 1969 pp. 324-338, https://www.researchgate.net/publication/23919490\_Space-time\_structure\_in\_high\_energy\_interactions

$\left[  23\right]  $ Bennett, C H.: The Thermodynamics of Computation -- a
Review. Int. J. of Theor, Phys 21, 905-940 (1982)

$\left[  24\right]  $ Bennett, C. H., Bernstein, E., Brassard, G., and
Vazirani, U.: Strengths and Weaknesses of Quantum Computing. SIAM Journal on
Computing Vol. 26 Issue 5, 1510-1523 (1997)

$\left[  25\right]  $ Long, G. L.: Grover algorithm with zero theoretical
failure rate. Phys. Rev. A 64, 022307-022314 (2001)

$\left[  26\right]  $ Toyama, F. M., van Dijk, W., and Nogami, Y.: Quantum
search with certainty based on modified Grover algorithms: optimum choice of
parameters. Quantum Information Processing 12, 1897-1914 (2013)

$\left[  27\right]  $ Feynman R. and Hibbs A. R. "Quantum Mechanics And Path
Integrals" New York McGraw-Hill (1965).

$\left[  28\right]  $\ Deutsch, D. and Jozsa, R.: Rapid solution of problems
by quantum computation. Proc. Roy. Soc. A 439, 553-558 (1992)

$\left[  29\right]  $ Simon, D.: On the power of quantum computation.
Proceedings of the 35th Annual IEEE Symposium on the Foundations of Computer
Science 116-123 (1994)

$\left[  30\right]  $ Kaye, P., Laflamme, R., and Mosca, M.: An Introduction
To Quantum Computing. Oxford University Press 146-147 (2007)

$\left[  31\right]  $ Ollivier, H. and Zurek, W. H.: Quantum Discord: A
Measure of the Quantumness of Correlations. Phys. Rev. Lett. 88, 017901-017909 (2001)

$\left[  32\right]  $ Henderson, L. and Vedral, V.: Classical, quantum and
total correlations. Journal of Physics A 34, 6899- 6709 (2001)

$\left[  33\right]  $ Jozsa, R. and Linden, N.: On the role of entanglement in
quantum computational speed-up. Proc. Roy. Soc. A 1097 (2002)

$\left[  34\right]  $ Gross, D., Flammia, S. T., and Eisert, J.: Most quantum
states are too entangled to be useful as computational resources. Phys. Rev.
Lett. 102, 190501: 1-4 (2009)

$\left[  35\right]  $ Aaronson, S.: Quantum computing, postselection, and
probabilistic polynomial-time. Proc. Roy. Soc. A 461, 3473-3482 (2005)

$\left[  36\right]  $ Cai, Y., Le, H. N., and Scarani, V.: State complexity
and quantum computation. Annalen der Physik 527 (9) (March 2015)

$\left[  37\right]  $ Howard, M., Wallman, J., Veitch, V., and Emerson, J.:
Contextuality supplies the `magic' for quantum computation. Nature 510,
351--355 (2014)

$\left[  38\right]  $ Morikoshi, F.: Information-theoretic temporal Bell
inequality and quantum computation. Phys. Rev. A 73, 052308-052312 (2006)

$\left[  39\right]  $ Price, H.: Locality, Independence and the Pro-Liberty
Bell. 10th International Symposium of Logic, Methodology and Philosophy of
Science, Florence. arxiv.quant-ph/9602020 (1995)

$\left[  40\right]  $ Price, H. and Wharton, K.: Disentangling the Quantum
World. arxiv.org/abs/1508.01140 (2015)

$\left[  41\right]  $ Grover, L. K.: From Sch\"{o}dinger's Equation to the
Quantum Search Algorithm. arXiv:quant-ph/0109116 (2001)

$\left[  42\right]  $ Price, H. and Farr, M.:\ Program of the Conference: Free
will and retrocausality in the quantum world. Trinity College, Cambridge (2014)

$\left[  43\right]  $ Finkelstein, D. R.: Private communication (2010)

$\left[  44\right]  $ Eitzur, A. C.: Private communication (2016)

\end{document}